\documentclass[a4paper]{jpconf}
\usepackage{graphicx}
\usepackage{lineno}

\begin{document}

%\setpagewiselinenumbers  
%\modulolinenumbers[5]
%\linenumbers

\title{Ligth-flavour identified charged-hadron production in pp and Pb--Pb collisions at the LHC}

\author{Roberto Preghenella for the ALICE Collaboration}

\address{Centro Studi e Ricerche e Museo Storico della Fisica ``Enrico Fermi'', Rome, Italy \\Sezione INFN, Bologna, Italy}

\ead{preghenella@bo.infn.it}

\begin{abstract}

Thanks to the unique detector design adopted to fulfill tracking and particle-identification (PID) requirements (e.g. low momentum cut-off and low material budget), the ALICE experiment provides significant information about hadron production both in pp and Pb--Pb collisions. In particular, the $p_{\rm T}$-differential and integrated production yields of identified particles play a key role in the study of the collective and thermal properties of the matter formed in high-energy heavy-ion collisions. Furthermore, the production of high-$p_{\rm T}$ particles provides insights into the property of the hot medium created in such collisions and the in-medium energy-loss mechanisms.

Transverse momentum spectra of $\pi^{\pm}$, K$^{\pm}$, p and $\bar{\rm p}$ are measured at mid-rapidity ($\left|y\right|~<~0.5$) over a wide momentum range, from $\sim$~100 MeV/$c$ up to $\sim$~20 GeV/$c$. The measurements are performed exploiting the d$E$/d$x$ in silicon and gas, the time-of-flight and the ring-imaging Cherenkov particle-identification techniques, which will be briefly reviewed in this report. The current results on light-flavour charged-hadron production will be presented for pp collisions at $\sqrt{s}$ = 0.9, 2.76 and 7 TeV and for Pb--Pb collisions at $\sqrt{s_{\rm NN}}$ = 2.76 TeV. Integrated production yields, transverse momentum spectra and particle ratios in pp are discussed as a function of the collision energy and compared to previous experiments and commonly-used Monte Carlo models. Pb--Pb collisions at the LHC feature the highest radial flow ever observed and an unexpectedly low p/$\pi$ production ratio. The results are presented as a function of collision centrality and compared to RHIC data in Au--Au collisions at $\sqrt{s_{\rm NN}}$ = 200 GeV and predictions from thermal and hydrodynamic models. The nuclear modification factor ($R_{\rm AA}$) of identified hadrons will also be discussed and compared to unidentified charged particles and theoretical predictions. This is observed to be identical for all particle species at high-$p_{\rm T}$.
\end{abstract}

\section{Introduction}

ALICE (A Large Ion Collider Experiment) is a general-purpose heavy-ion detector at the CERN LHC (Large Hadron Collider). It has been designed in order to fulfill the requirements to track and identify particles from very low ($\sim$100~MeV/$c$) up to quite high ($\sim$100~GeV/$c$) transverse momenta in an environment with large charged-particle multiplicities as in the case of central lead-lead (Pb--Pb) collisions at the LHC.

\begin{figure}[t]
\centering
\includegraphics[width=0.7\linewidth,clip]{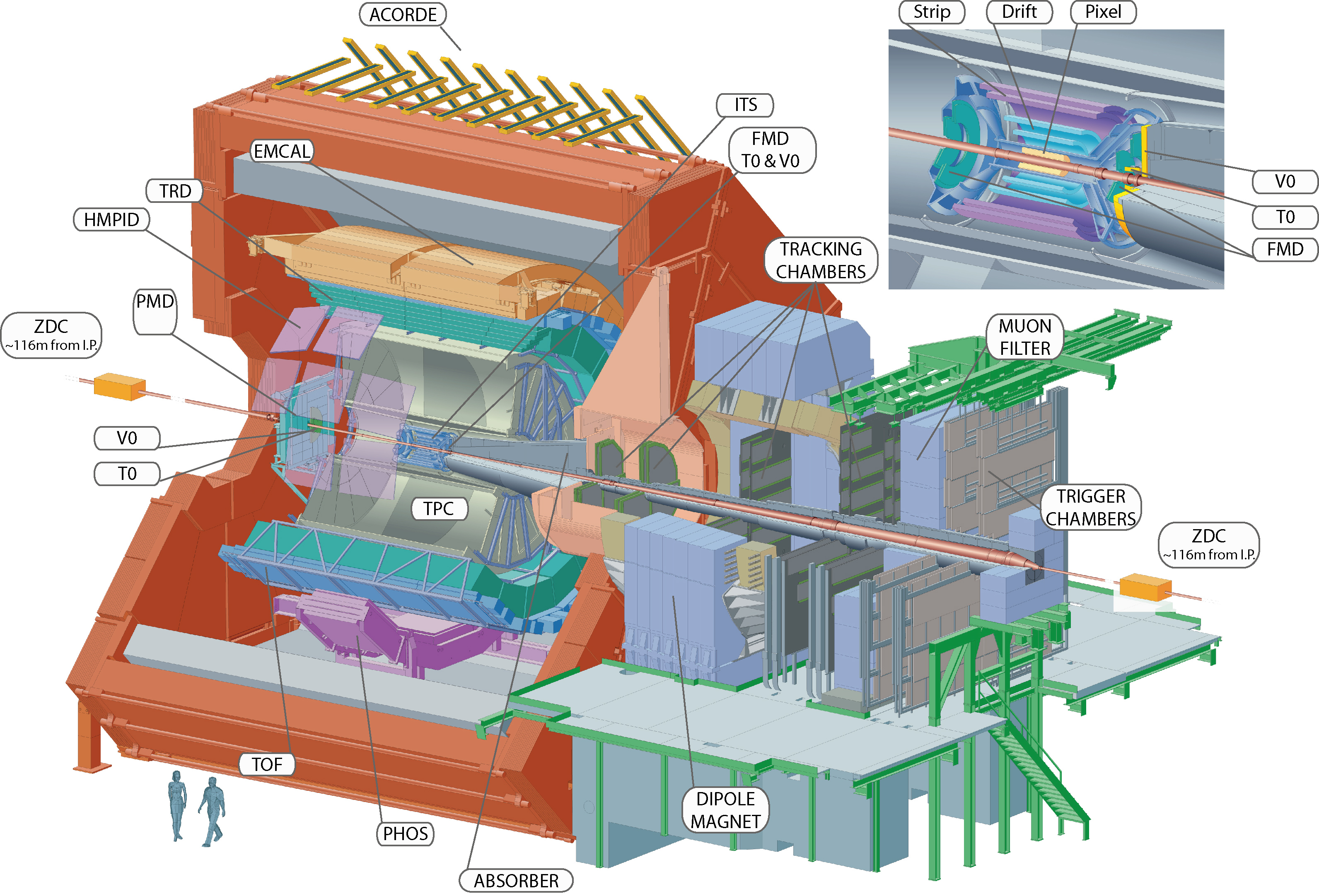};
\caption{Schematic layout of the ALICE detector with its main subsystems.}
\label{fig:alice}
\end{figure}

The ALICE experiment, shown in Figure~\ref{fig:alice}, consists of a central-barrel detector and several forward detector systems.
The central system covers the mid-rapidity region ($\left| \eta \right| \leq$ 0.9) over the full azimuthal angle. It is installed inside a large solenoidal magnet providing a moderate magnetic field of 0.5 T. It includes a six-layer high-resolution inner-tracking system (ITS), a large-volume time-projection chamber (TPC) and electron and charged-hadron identification detectors which exploit transition-radiation (TRD) and time-of-flight (TOF) techniques, respectively. Small-area systems for high-$p_{\rm T}$ particle-identification (HMPID), photon and neutral-meson measurements (PHOS) and jet reconstruction (EMCal) complement the central barrel. Thanks to these unique features the experiment
is able to identify hadrons in a wide momentum range by employing
different detection systems and techniques, as discussed in 
Section~\ref{sec:pid}.
The detectors which cover larger rapidity regions include a single-arm muon spectrometer covering the pseudorapidity range -4.0 $\leq \eta \leq$ -2.4 and several smaller detectors (VZERO, TZERO, FMD, ZDC, and PMD) for triggering, multiplicity measurements and centrality determination. A detailed description of the ALICE detector layout and of its subsystems can be found in~\cite{Aamodt:2008zz}.

Since November 2009 when the first collisions at the LHC occurred, the ALICE experiment has collected proton-proton data at several centre-of-mass energies ($\sqrt{s}$ = 0.9, 2.76, 7 and 8 TeV). During the first two LHC heavy-ion runs, in 2010 and 2011, the detector recorded Pb--Pb collisions at a centre-of-mass energy per nucleon pair of $\sqrt{s_{\rm NN}}$ = 2.76 TeV and could profit from of an integrated luminosity of about 10 $\mu \rm b^{-1}$ and 100 $\mu \rm b^{-1}$, respectively. It is worth stressing the outstanding performance of the LHC complex: the instant luminosity exceeded $10^{26}\rm cm^{-2} s^{-1}$ in the second run, higher than the design value. 
Proton-lead (p--Pb) collision data were also collected with the ALICE detector. This occurred during a short run performed in September 2012 in preparation for the main p--Pb run at the beginning of 2013. Eight pairs of bunches collided in the ALICE interaction region, providing a luminosity of about $8 \times 10^{25}\rm cm^{-2}s^{-1}$. The configuration (4 TeV protons colliding with fully stripped $^{208}$Pb ions at $82 \times 4$ TeV) resulted in interactions at $\sqrt{s_{\rm NN}}$ = 5.02 TeV in the nucleon-nucleon centre-of-mass system, which moves with a rapidity of $\Delta y_{\rm NN}$ = 0.465.

\section{Particle identification}\label{sec:pid}

\begin{figure}[t]
\centering
\includegraphics[width=0.49\linewidth,clip]{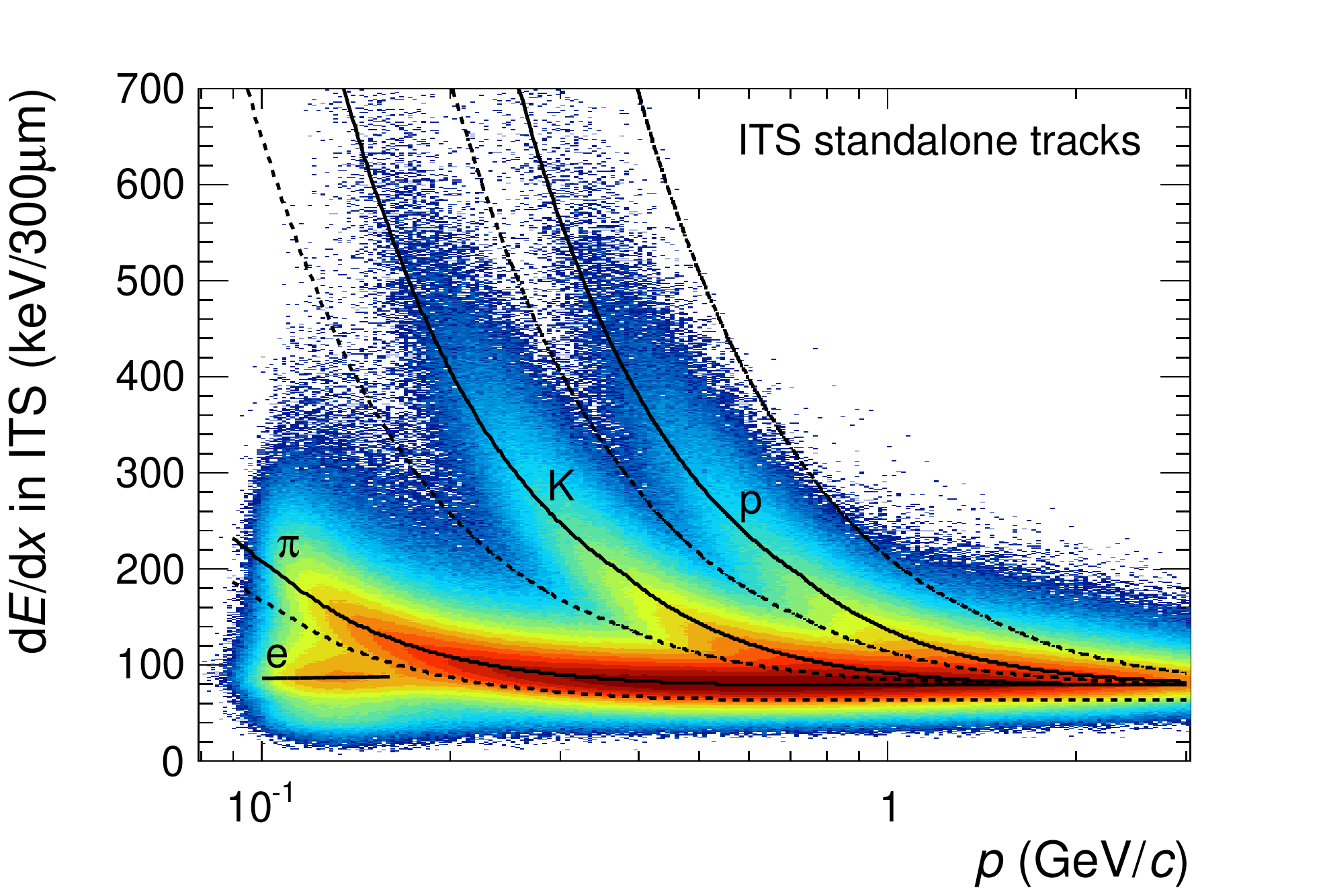}
\includegraphics[width=0.49\linewidth,clip]{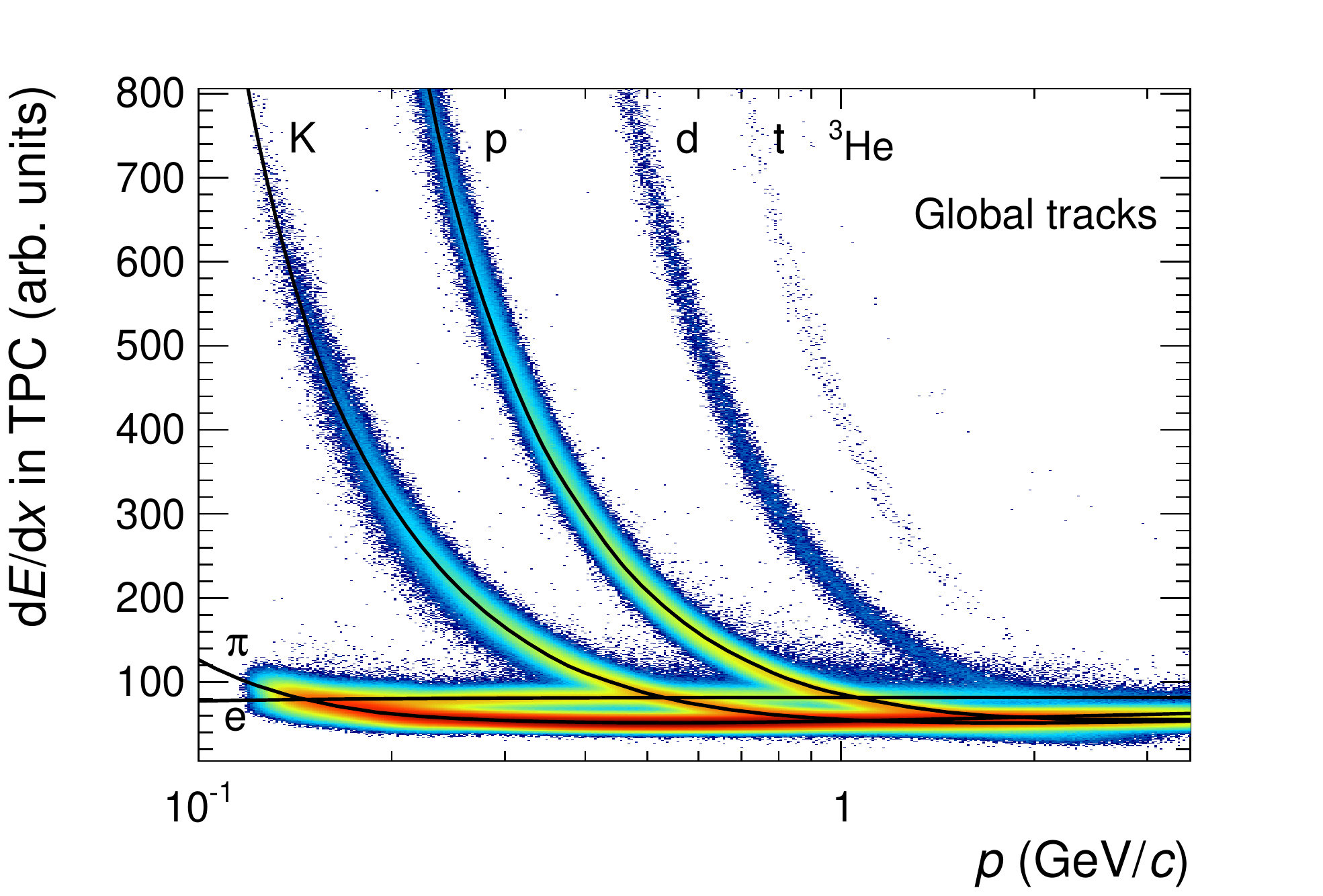}
\includegraphics[width=0.49\linewidth,clip]{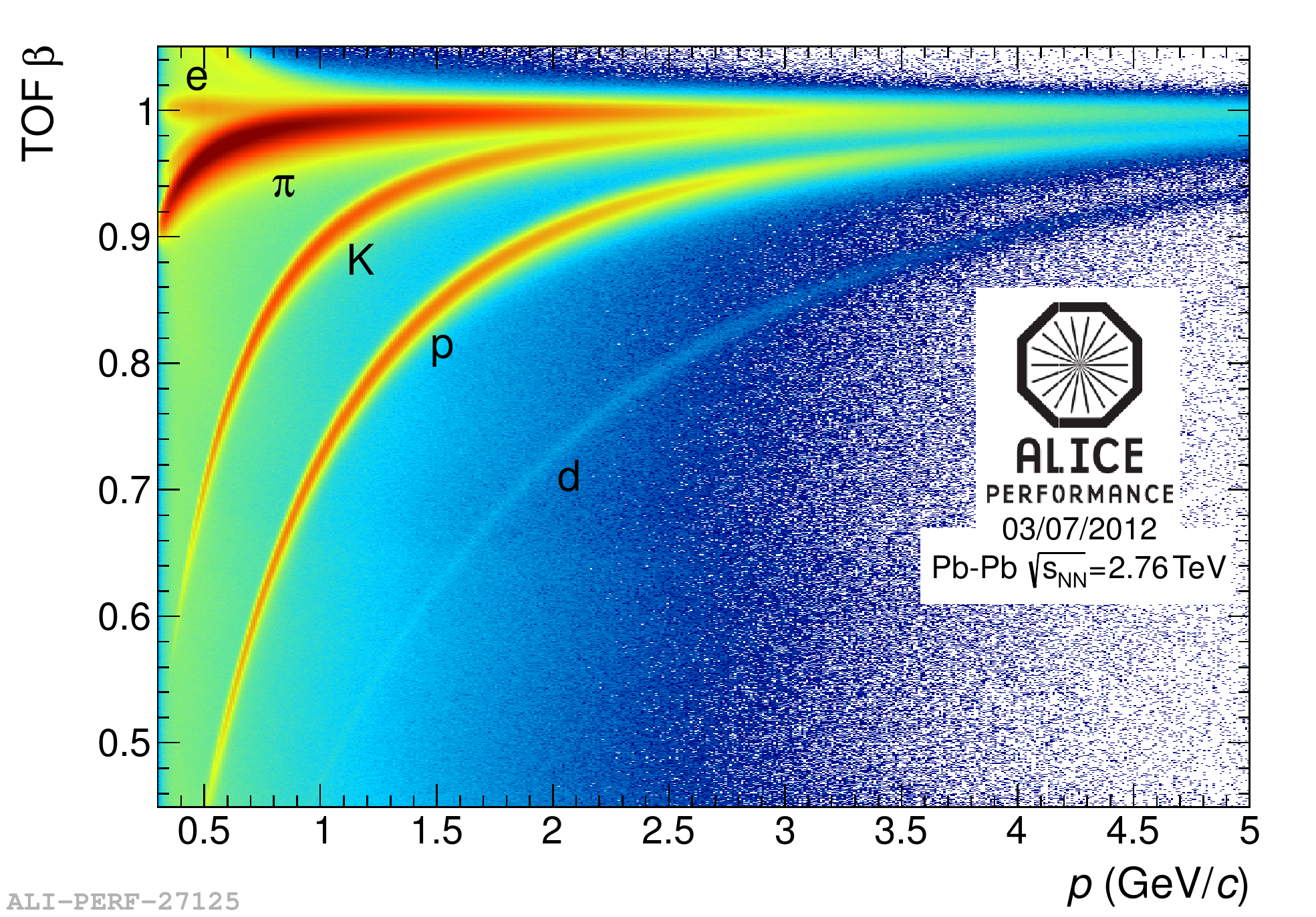}
\includegraphics[width=0.49\linewidth,clip]{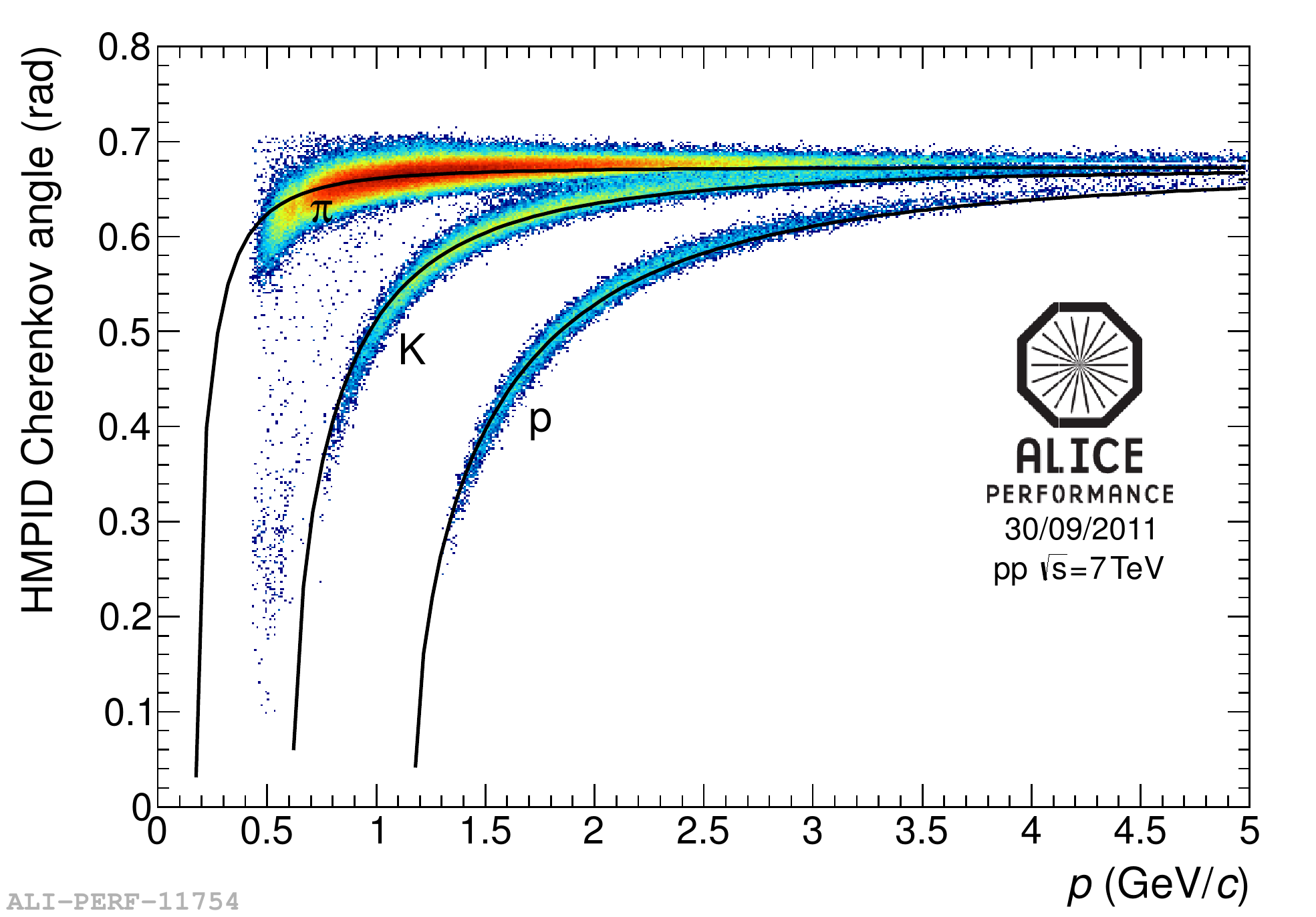}
\caption{Energy loss d$E$/d$x$ in the ITS (top-left) and in the TPC (top-right). The continuous curves represent the Bethe-Bloch parametrization. (bottom-left) particle velocity $\beta$ measured with TOF as a function of momentum. (bottom-right) Cherenkov angle measured in the HMPID as a function of the track momentum.}
\label{fig:pid}
\end{figure}

\begin{figure}[t]
\centering
\includegraphics[width=0.6\linewidth,clip]{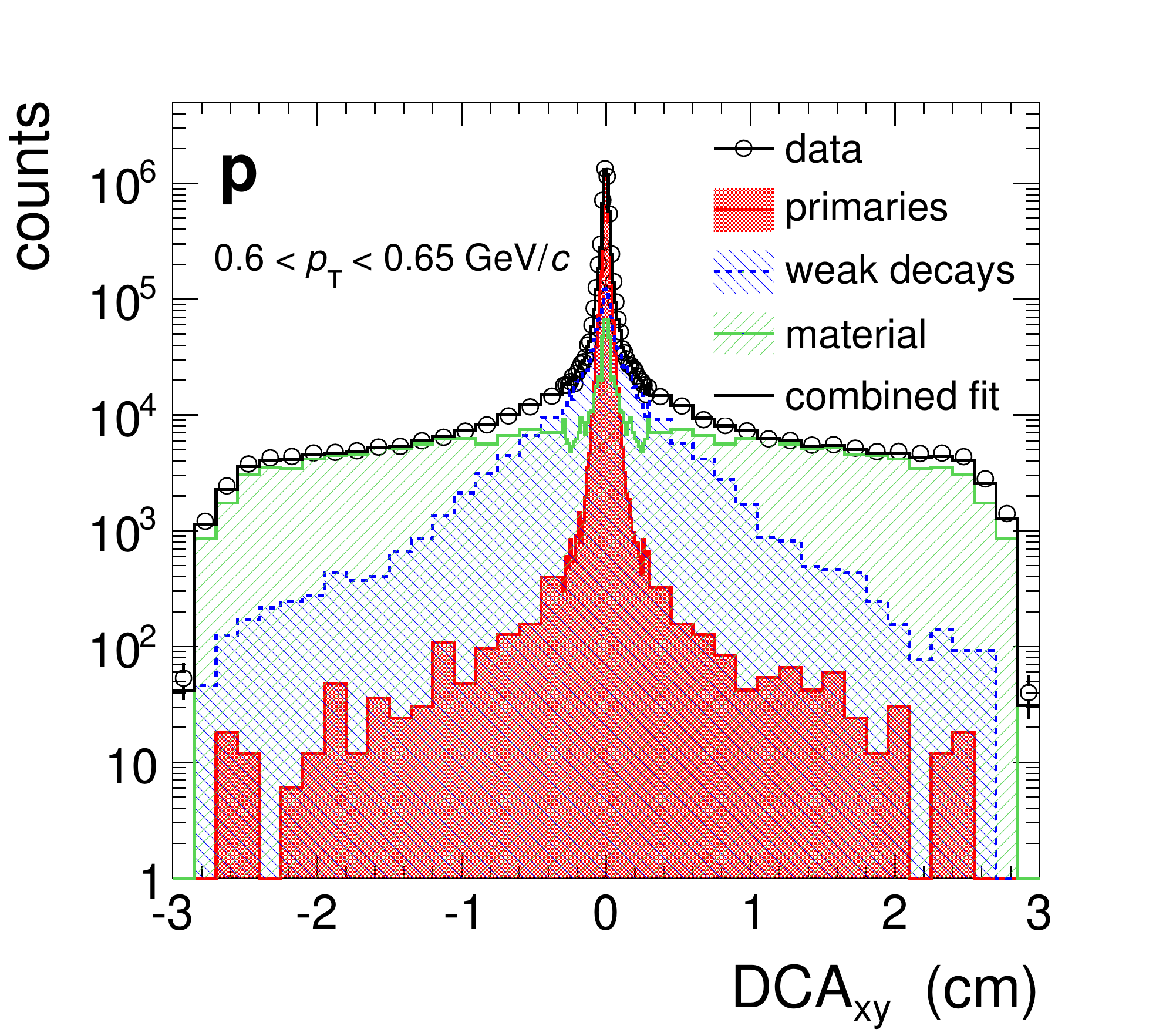}
\caption{Transverse DCA ($DCA_{xy}$) of protons in the range between 0.6 GeV/$c$ and 0.65 GeV/$c$ in 0-5\% most central Pb--Pb collisions together with the Monte Carlo templates which are fitted to the data.}
\label{fig:dca}
\end{figure}

In this section the main particle-identification (PID) detectors relevant to the
analyses presented in this paper are briefly discussed. A detailed review of the ALICE experiment and of its PID capabilities can
be found in~\cite{Alessandro:2006yt}. The ITS is the innermost detector system, a six-layer silicon detector located at radii between 4 and 43
cm. Four of the six layers provide specific energy loss d$E$/d$x$ measurements and are used for
particle identification in the non-relativistic ($1/\beta^2$)
region. By using the ITS as a standalone tracker it is possible to
reconstruct and identify low-momentum particles (below~200~MeV/c) not reaching the main tracking
systems (Figure~\ref{fig:pid} top-left). The TPC is the main central-barrel tracking detector of ALICE. It
provides three-dimensional hit information and specific energy-loss
measurements with up to 159 samples. With the measured particle momentum and
$\langle $d$E$/d$x \rangle$ the particle type can be determined by comparing the
measurements with the Bethe-Bloch expectation (Figure~\ref{fig:pid} top-right). The TOF detector is a
large-area array of Multigap Resistive Plate Chambers (MRPC) and covers the central
pseudorapidity region ($\left| \eta \right| <$~0.9, full azimuth). Particle
identification is performed by matching momentum and trajectory-length
measurements performed by the tracking system with the time-of-flight
information provided by the TOF system. The overall time-of-flight resolution is
measured to be about 85~ps in Pb--Pb collisions (about 120~ps in pp collisions) and it is determined by the time resolution
of the detector itself and by the start-time resolution (Figure~\ref{fig:pid} bottom-left). The HMPID detector consists of seven identical proximity focusing RICH (Ring Imaging Cherenkov) counters. Photon detection is performed using the proportional multiwire chambers coupled to pad-segmented CsI photocathode. Particle identification is obtained by means of the measurement of the Cherenkov angle allowing the separation of pions and kaons between 1 GeV/$c$ and 3 GeV/$c$ and protons from 1.5 GeV/$c$ up to 5 GeV/$c$ (Figure~\ref{fig:pid}).

The transverse momentum spectra of primary $\rm \pi^{\pm}$, $\rm K^{\pm}$,
$\rm p$ and
$\rm \bar{p}$ are measured at mid-rapidity ($\left|y\right|~<~0.5$) combining the
techniques and detectors described above. Primary particles
are defined as prompt particles produced in the collision and all decay daughters, except products from weak decays of strange particles. The
contribution from the feed-down of weakly-decaying particles to $\rm \pi^{\pm}$,
$\rm p$ and $\rm \bar{p}$ and from protons emitted from secondary interactions with material are subtracted by fitting the
data using Monte Carlo templates of the DCA\footnote{Distance of Closest
  Approach to the reconstructed primary vertex.} distributions (Figure~\ref{fig:dca}). Particles can
also be identified in ALICE via their characteristics decay 
topology or invariant mass. This, combined with the direct identification
of the decay daughters, allows one to reconstruct weakly-decaying particles and
hadronic resonances with a good signal-to-background ratio.

\section{Light-flavour hadron production}

\begin{figure}[t]
\centering
\includegraphics[width=0.49\linewidth,clip]{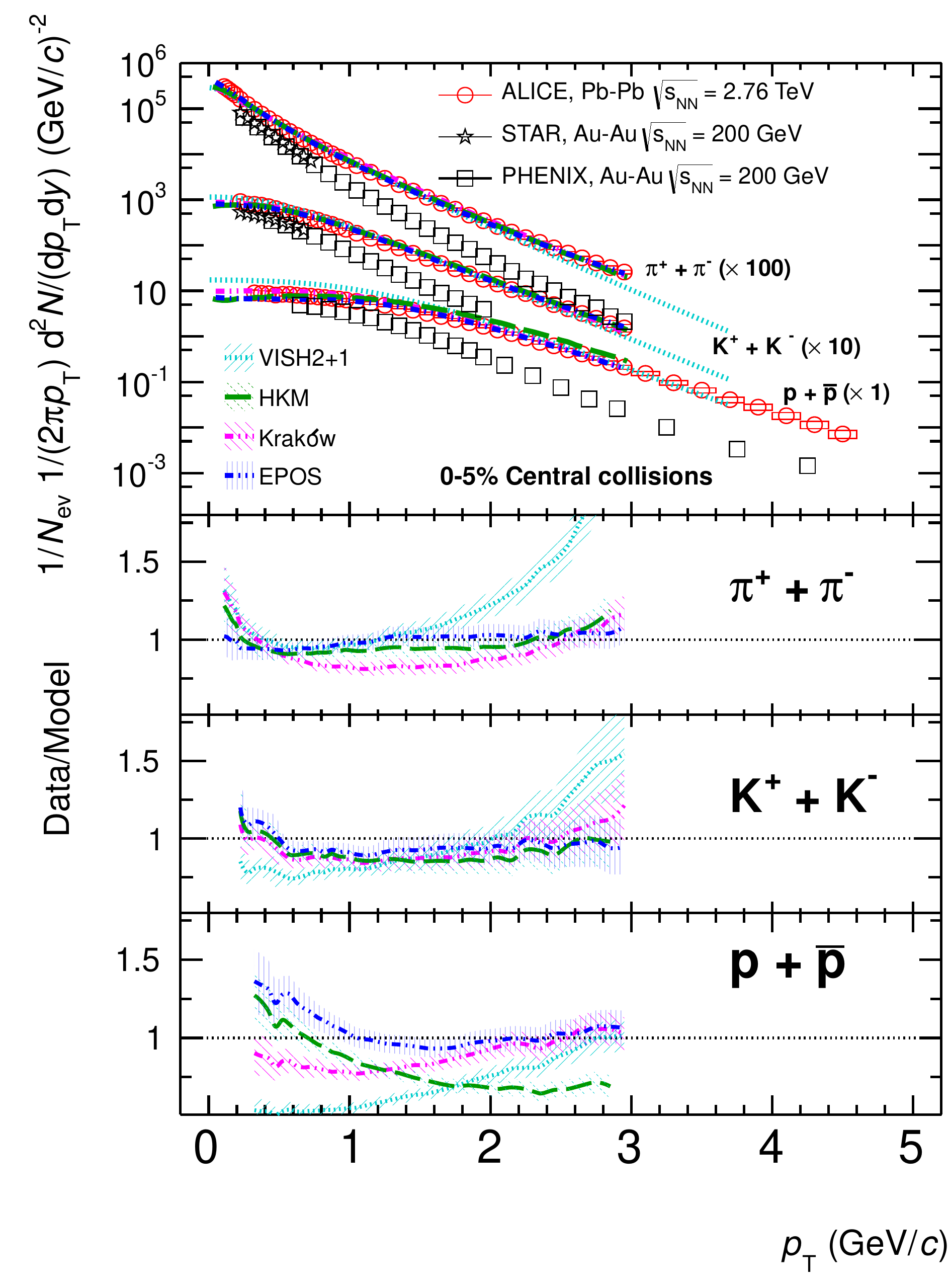}
\includegraphics[width=0.49\linewidth,clip]{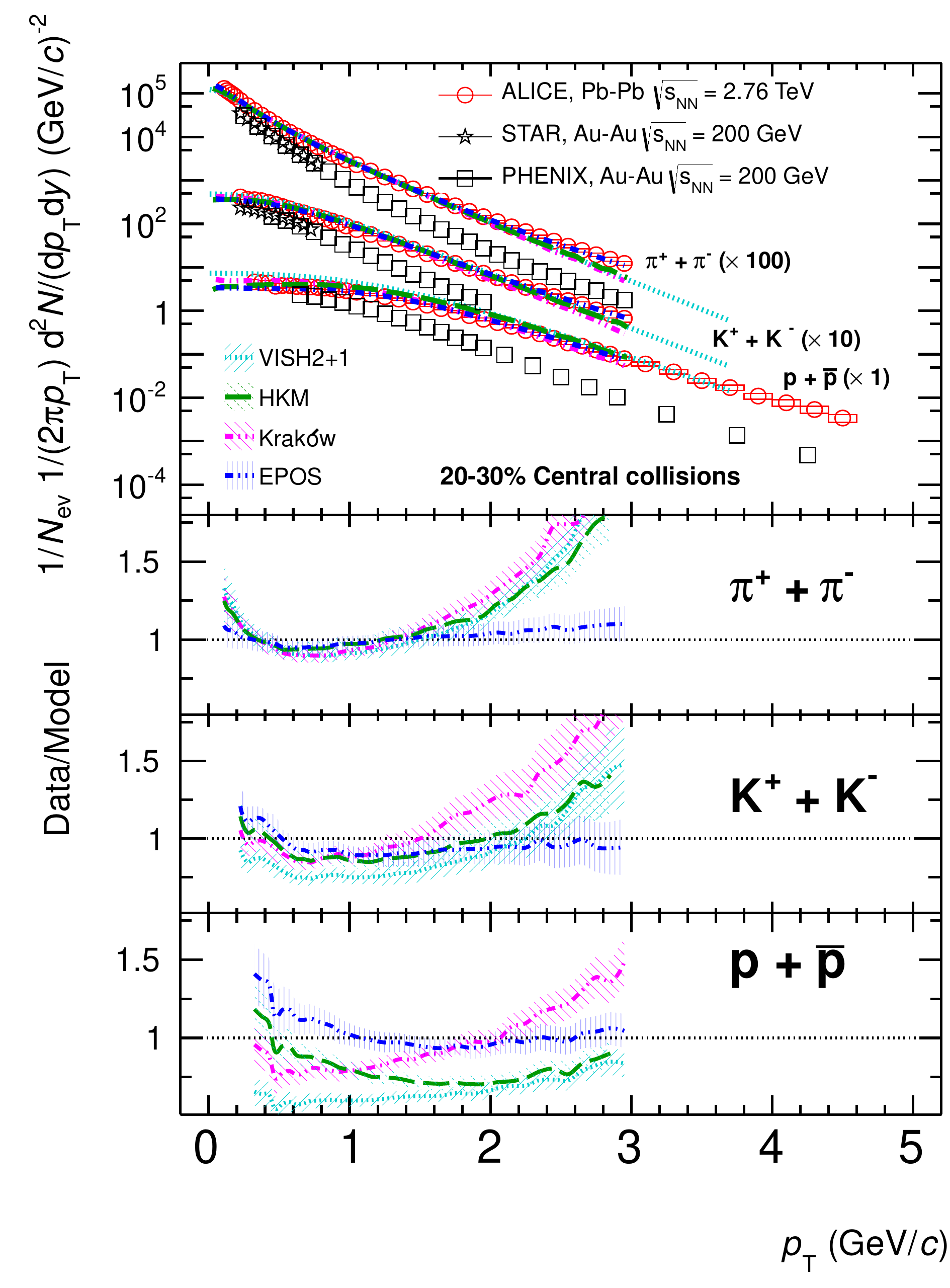}
\caption{Transverse momentum distributions of the sum of positive and negative pions, kaons and protons for central Pb--Pb collisions. The results are compared to RHIC data and hydrodynamic models.}
\label{fig:spectrapbpb}
\end{figure}

\begin{figure}[t]
\centering
\includegraphics[width=0.49\linewidth,clip]{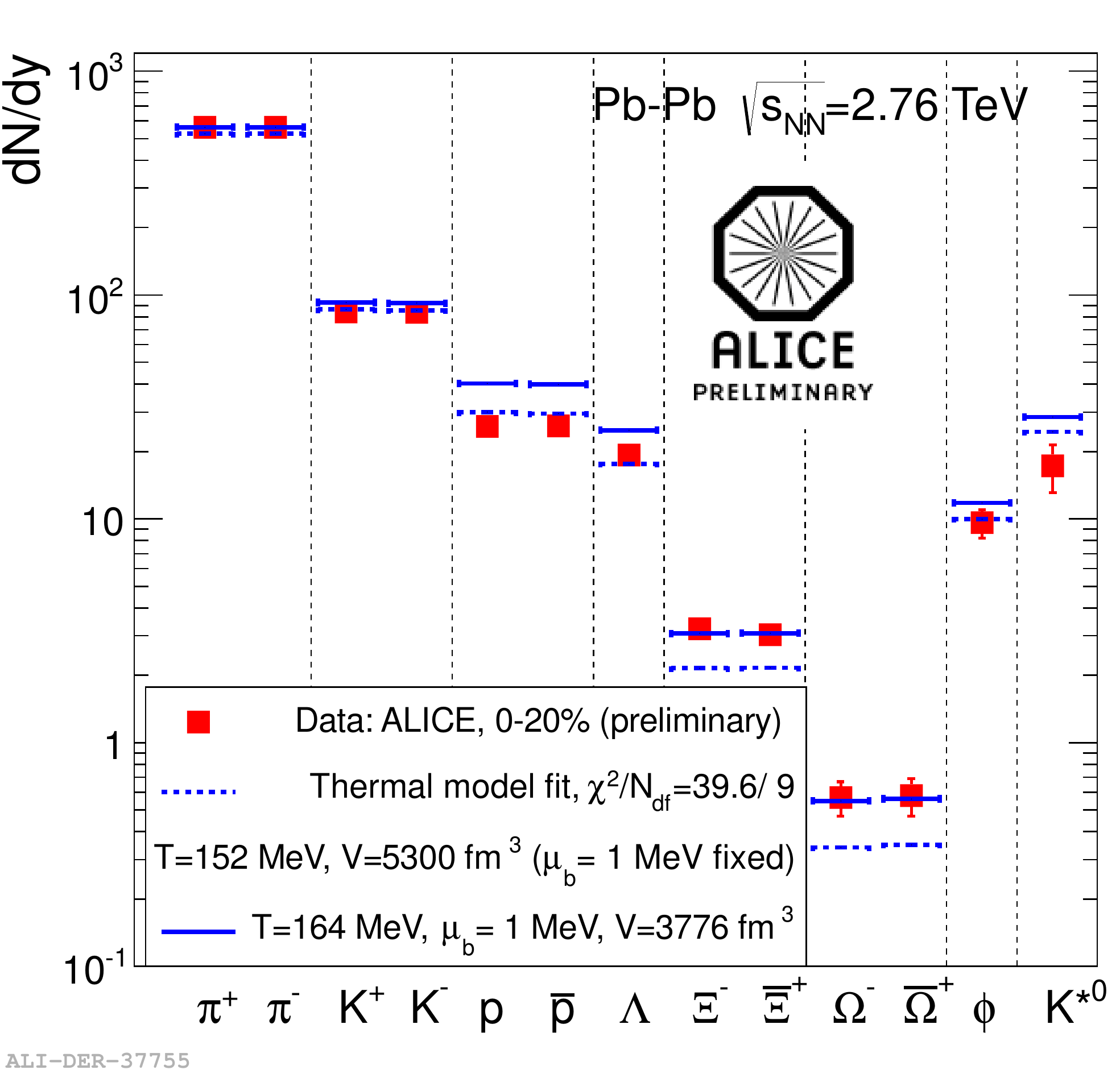}
\includegraphics[width=0.49\linewidth,clip]{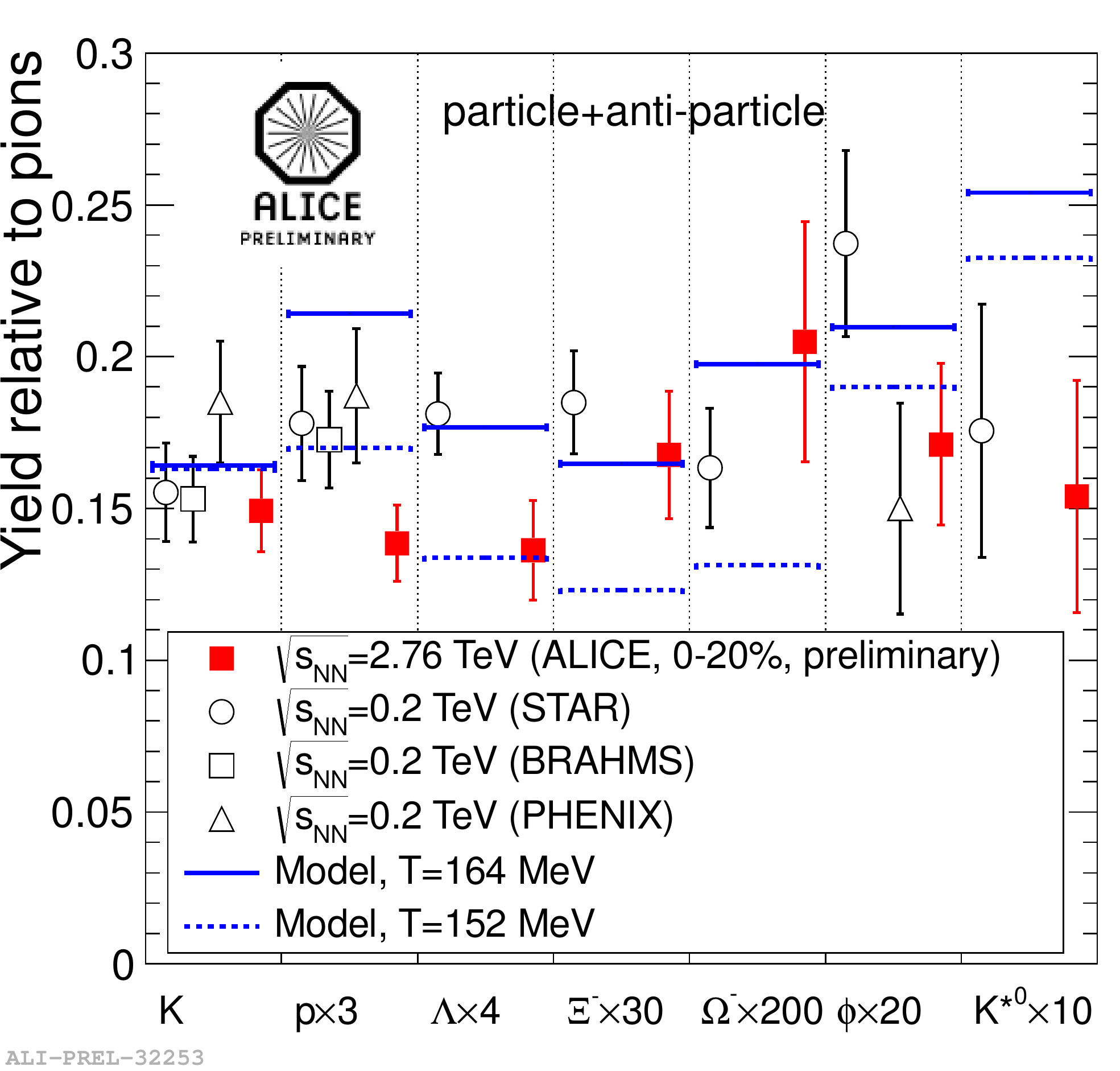}
\caption{(left) The thermal fit of ALICE data showing data and model for the best fit. (right) Mid-rapidity particle ratios compared to RHIC results and predictions from thermal models for central Pb--Pb collisions at the LHC.}
\label{fig:ratiospbpb}
\end{figure}

\begin{figure}[t]
\centering
\includegraphics[width=0.49\linewidth,clip]{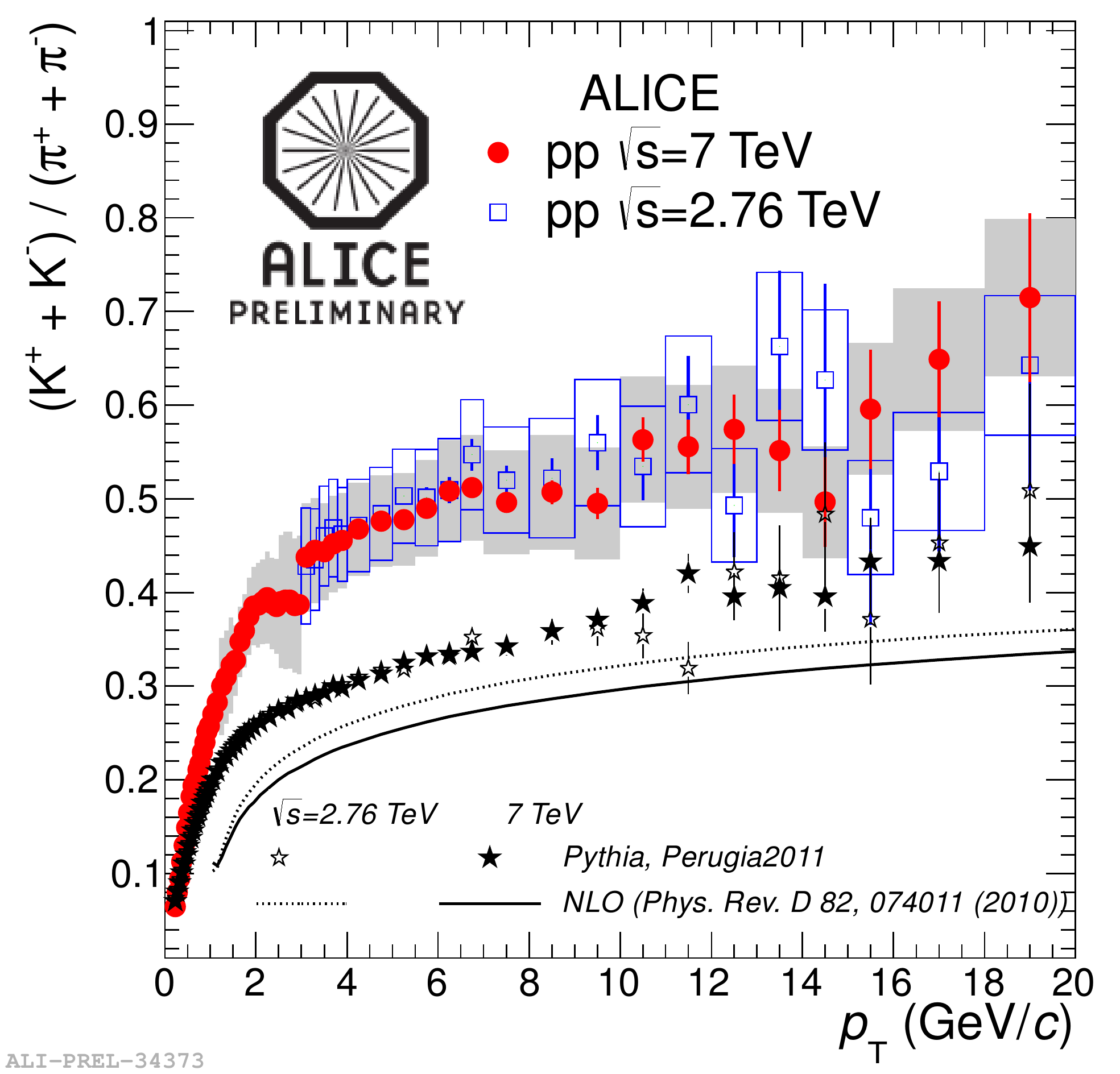}
\includegraphics[width=0.49\linewidth,clip]{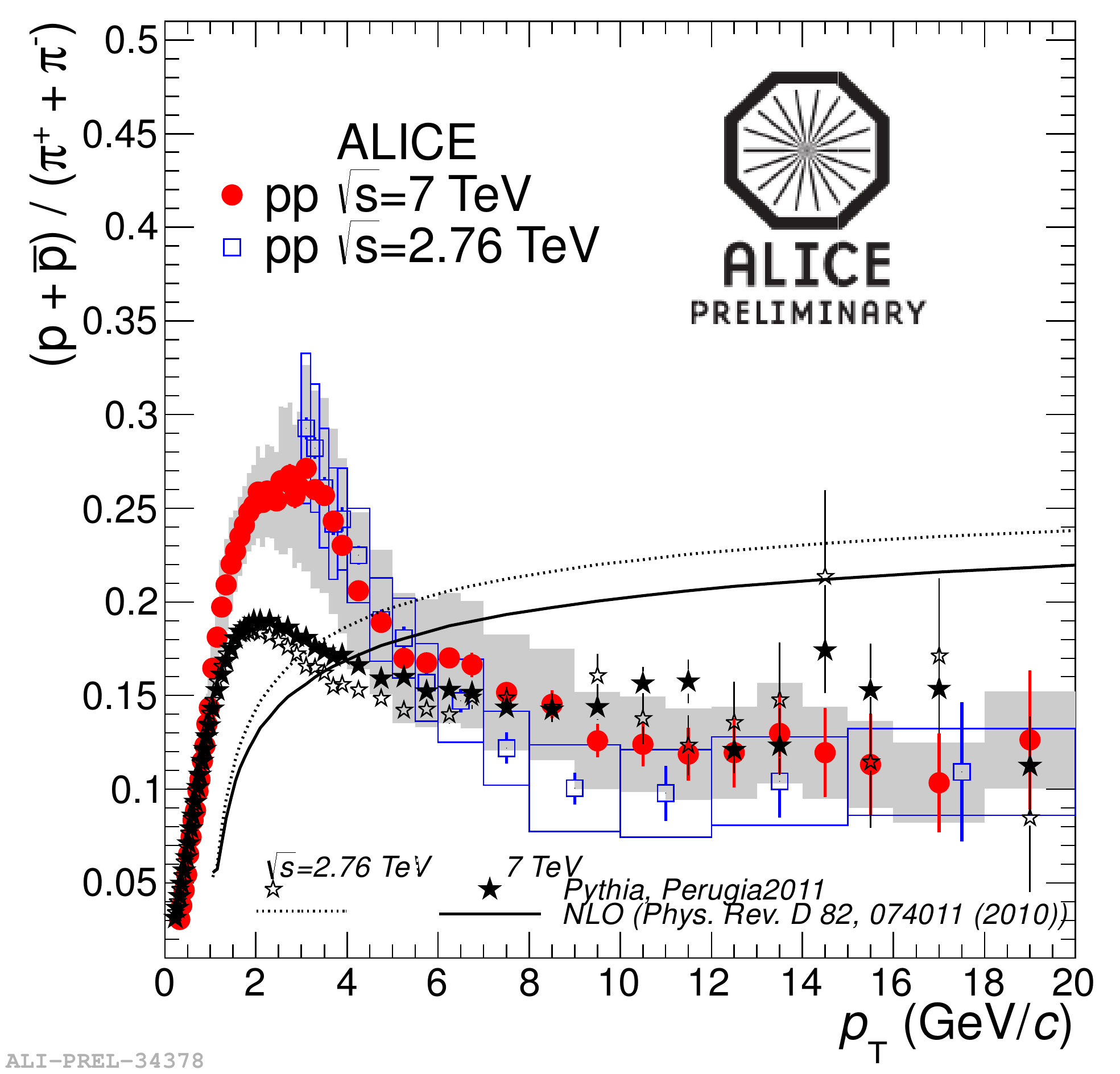}
\caption{$\rm K/\pi$ (top) and $\rm p/\pi$ production ratio in pp collisions compared with PYTHIA Monte Carlo predictions and NLO calculations.}
\label{fig:ratiospp}
\end{figure}

\begin{figure}[t]
\centering
\includegraphics[width=0.49\linewidth,clip]{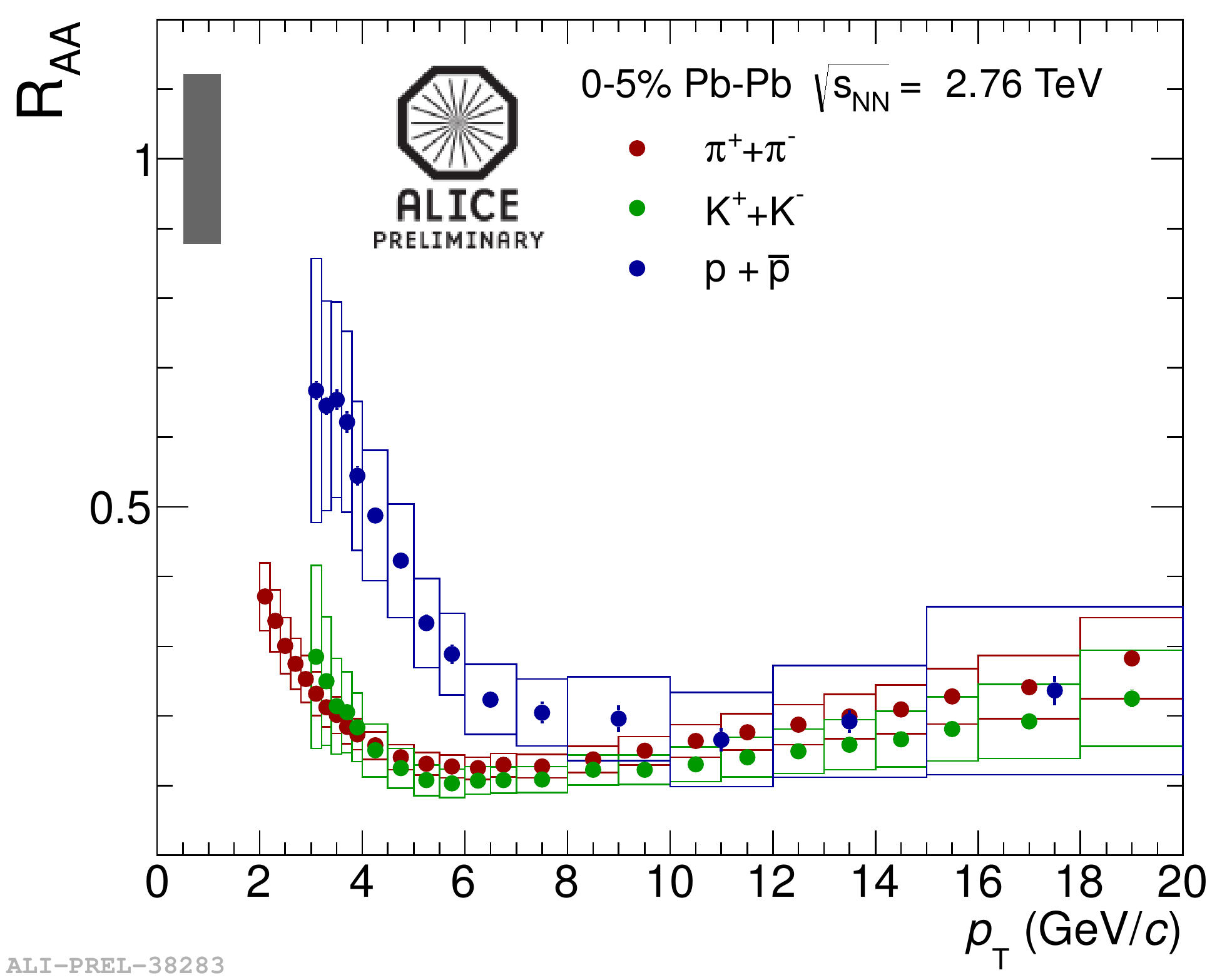}
\includegraphics[width=0.49\linewidth,clip]{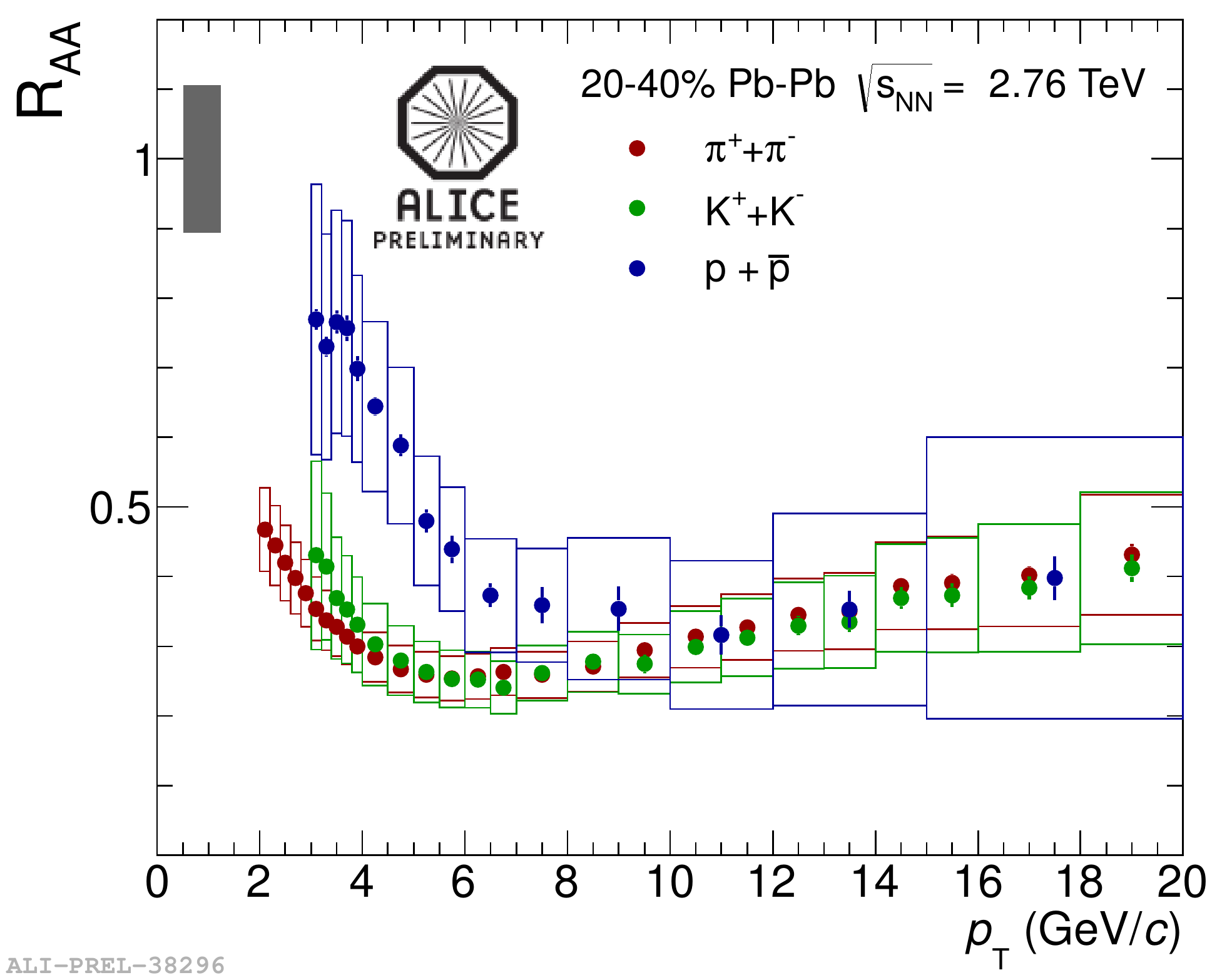}
\caption{Charged pion, kaon and (anti)proton nuclear modification factor $\rm R_{AA}$ as a function of $p_{\rm T}$ for Pb--Pb collisions at $\sqrt{s_{\rm NN}}$ = 2.76 TeV for centrality 0-5\% (left) and 20-40\% (right). Statistical (vertical error bars) and systematic (gray and colored boxes) are shown for the charged pion RAA. The gray boxes contains the common systematic error related to pp normalization to INEL and $\rm N_{coll}$.}
\label{fig:raa}
\end{figure}

\begin{figure}[t]
\centering
\includegraphics[width=\linewidth,clip]{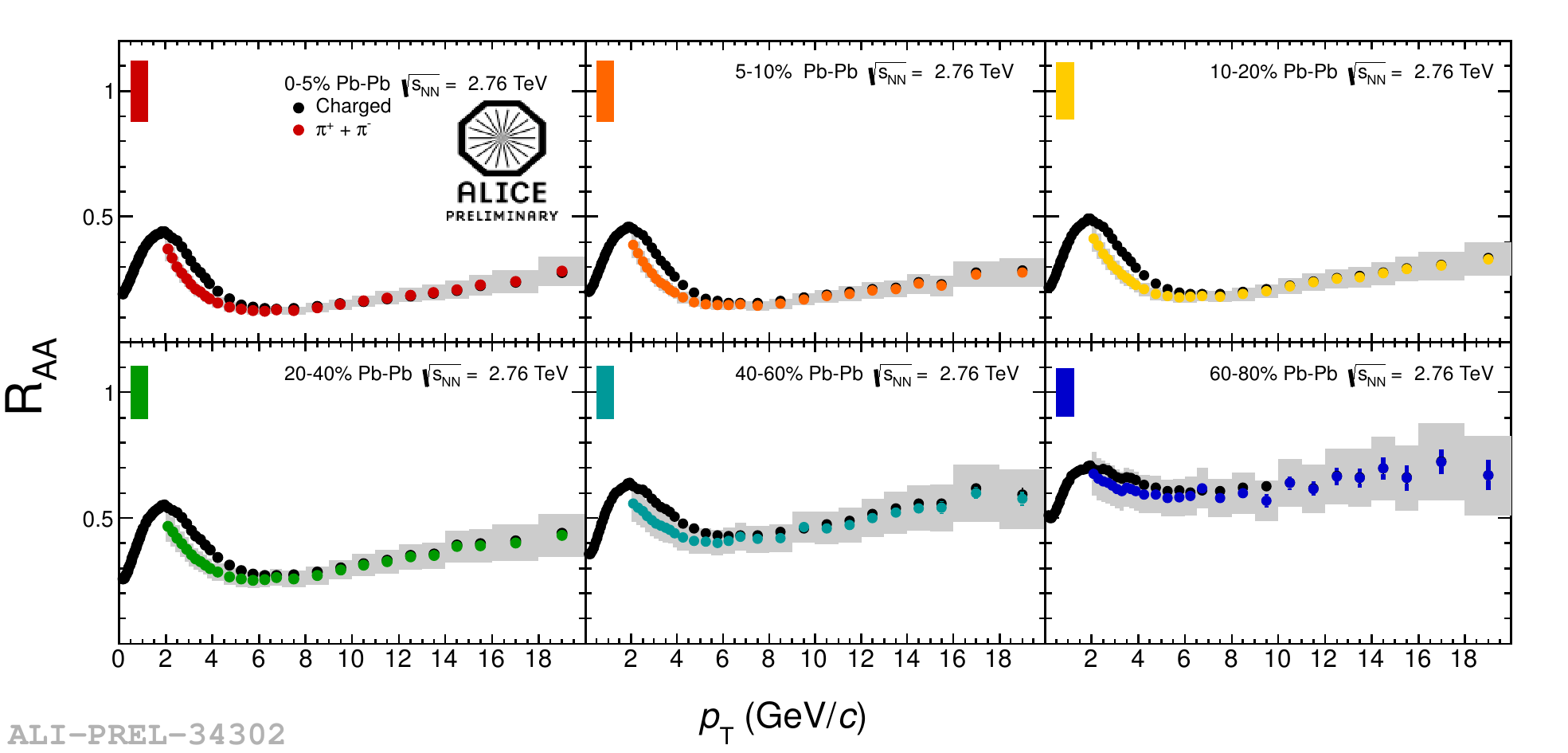}
\caption{Nuclear modification factor $\rm R_{AA}$ of charged pions compared to the $\rm R_{AA}$ of unidentified charged particles as a function of $p_{\rm T}$ for different centrality classes. Statistical (vertical error bars) and systematic (gray and colored boxes) errors are shown for the charged pions. The colored boxes contain the common systematic uncertainty related to the number of binary collisions and the pp normalization to INEL. Only statistical errors are shown for the unidentified charged $\rm R_{AA}$.}
\label{fig:pionraa}
\end{figure}

\begin{figure}[t]
\centering
\includegraphics[width=\linewidth,clip]{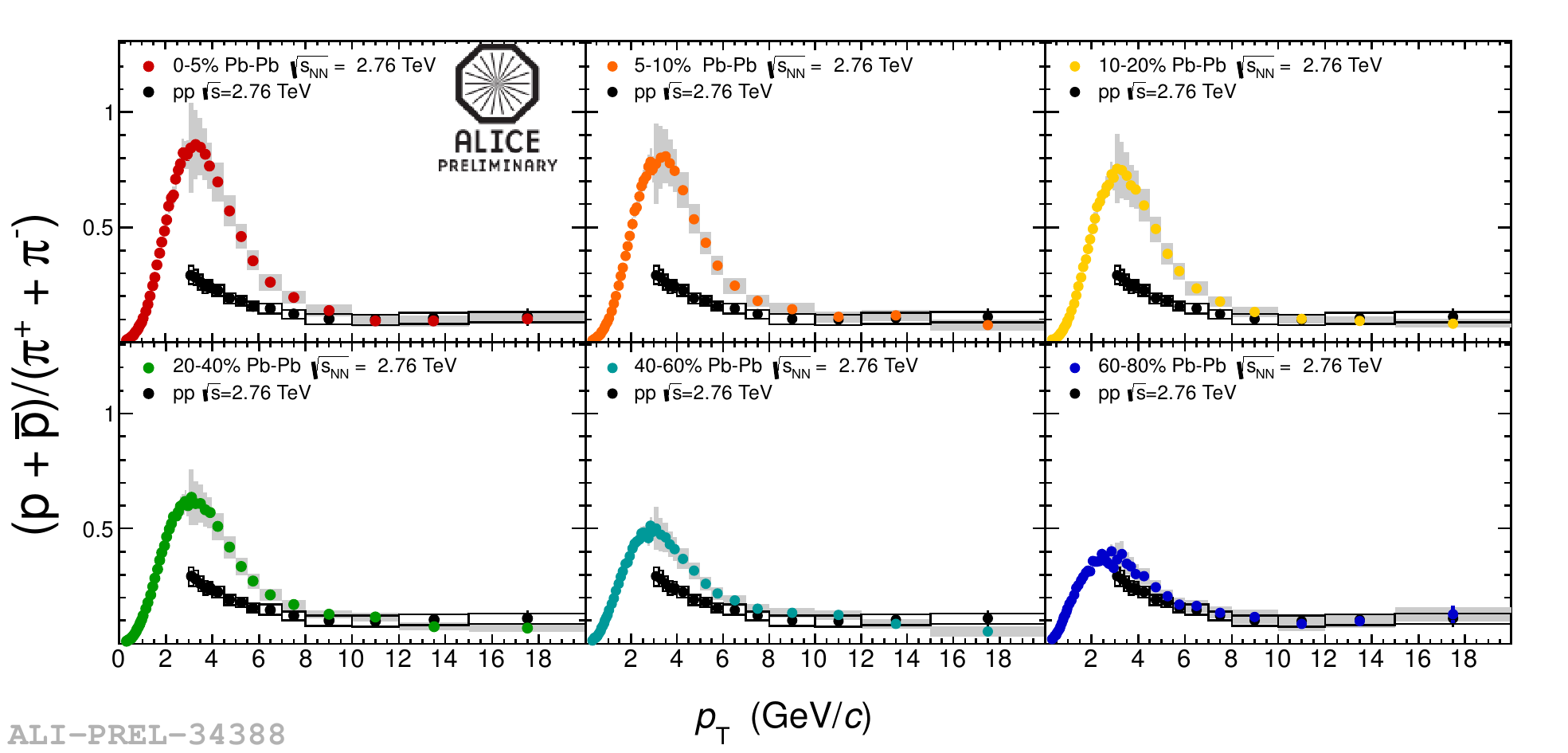}
\caption{Proton-to-pion ratio in Pb--Pb collisions compared with pp results at 2.76 TeV in several centrality bins. In Pb--Pb the statistical and systematic uncertainties are displayed as error bars and gray bands, respectively. In pp the systematic uncertainty is displayed in black squares.}
\label{fig:baryonmeson}
\end{figure}

\begin{figure}[t]
\centering
\includegraphics[width=0.49\linewidth,clip]{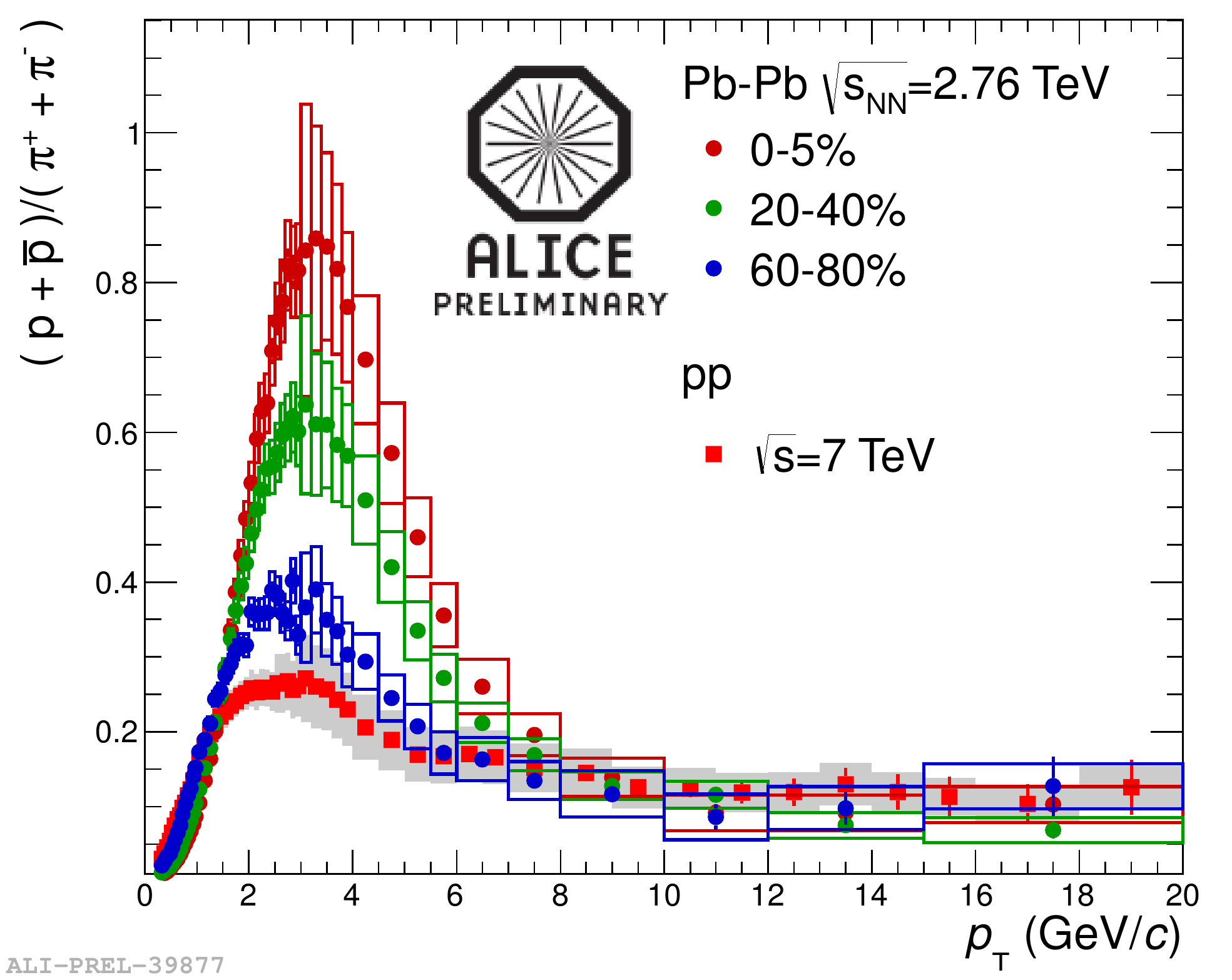}
\includegraphics[width=0.49\linewidth,clip]{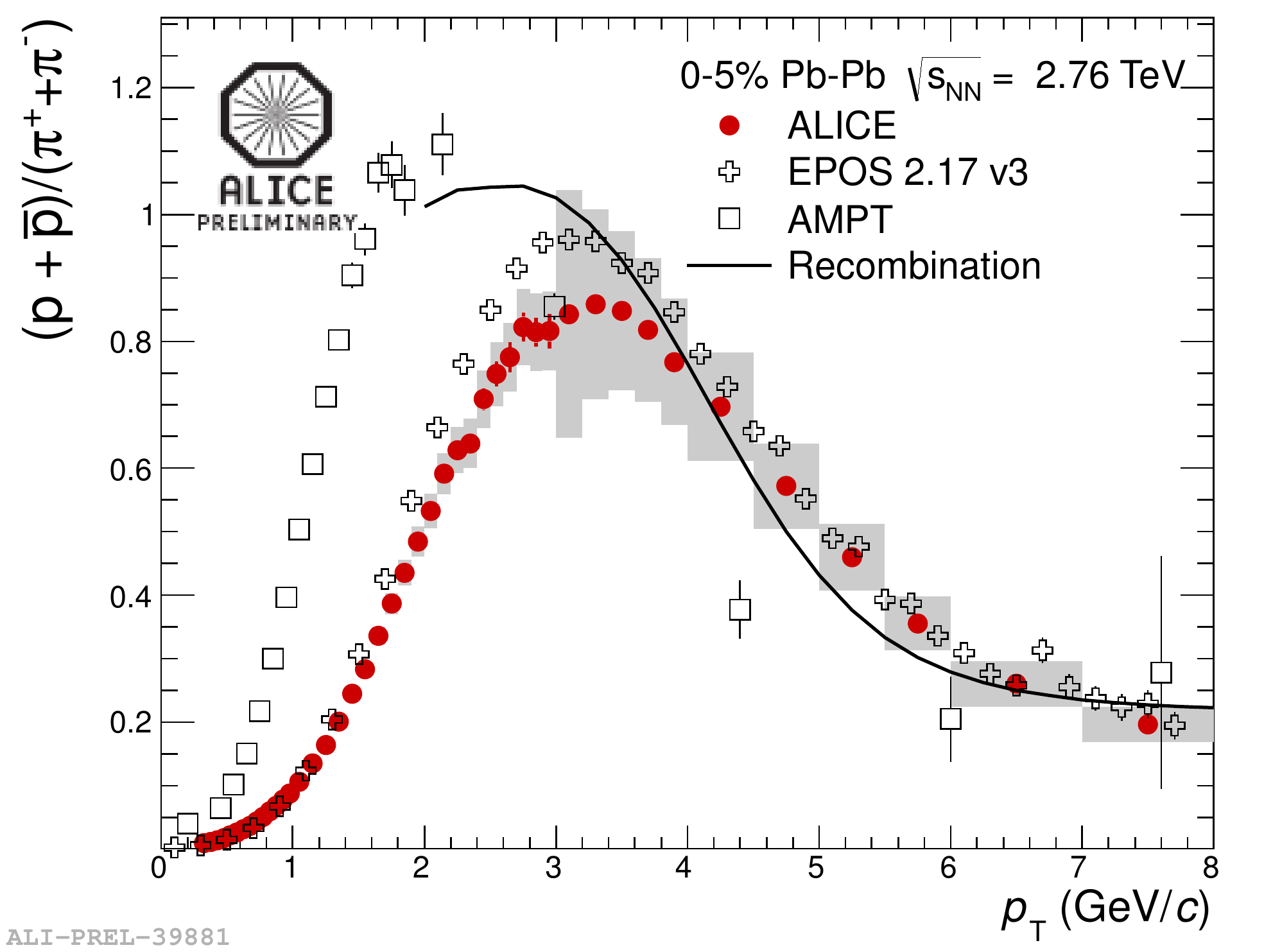}
\caption{(left) Proton-to-pion ratio in Pb--Pb collisions in several centrality bins compared with pp results at 2.76 TeV. (right) Proton-to-pion ratio in central Pb--Pb collisions compared with models.}
\label{fig:baryonmeson2}
\end{figure}

ALICE has measured the production yields of primary charged pions, kaon and (anti)protons in a wide momentum range and in several colliding systems. The measurements have been performed in proton-proton collisions at several centre-of-mass energies ($\sqrt{s}$ = 0.9, 2.76 and 7 TeV) and in Pb--Pb collisions at $\sqrt{s_{\rm NN}}$ = 2.76 TeV as a function of collision centrality.

In Pb--Pb collisions the transverse momentum $p_{\rm T}$ distributions and yields are compared to previous results at RHIC and expectations from hydrodynamic and thermal model. The results obtained for central Pb--Pb collisions are shown in Figure~\ref{fig:spectrapbpb} and~\ref{fig:ratiospbpb}. The spectral shapes are harder than those observed at RHIC, indicating an increase of the radial flow velocity with the centre-of-mass energy. The radial flow at the LHC is found to be about 10\% higher than at RHIC energy. 
The hydrodynamic models shown in Figure~\ref{fig:spectrapbpb} give for central collisions a fair description of the data, describing the experimental spectra within ∼20\%. This supports a hydrodynamic interpretation of the transverse momentum distribution in central collisions at the LHC. Further details of this anaysis and the models used to compare with the data can be found in~\cite{:2012iu}.

While the K/$\pi$ integrated production ratio is observed to be in line with lower energy measurements and predictions from the thermal model, both the p/$\pi$ and the $\Lambda/\pi$ ratios are lower than those at RHIC and significantly lower (a factor $\sim$ 1.5 and 1.35, respectively) than predictions. A possible explanation of these deviations from the thermal-model predictions may be re-interactions in the hadronic phase due to large cross sections for antibaryon-baryon annihilation~\cite{Steinheimer:2012rd,Becattini:2012sq,Pan:2012ne}.

The $p_{\rm T}$-dependent yield of charged kaons and protons normalized to charged pions are shown in Figure~\ref{fig:ratiospp} for pp collisions at $\sqrt{s}$ = 2.76 and 7 TeV where they are compared with theoretical model predictions. Within the experimental uncertainties no energy dependence is observed in the data. The observed production ratios are not reproduced by NLO calculations~\cite{Sassot:2010bh}. PYTHIA Monte Carlo generator~\cite{Skands:2010ak} underpredicts the proton-to-pion ratio at intermediate $p_{\rm T}$. 

Pion, kaon and (anti)proton production in Pb--Pb collisions have been compared to that in pp collisions and all show a suppression pattern which is similar to that of inclusive charged hadrons at high momenta ($p_{\rm T}$ above $\simeq$ 10 GeV/$c$), as shown in Figure~\ref{fig:raa} and~\ref{fig:pionraa}~\cite{OrtizVelasquez:2012te}. This seems to suggest that the dense medium formed in Pb--Pb collisions does not affect the fragmentation. A similar conclusion can be drawn from the proton-to-pion ratio measured in Pb--Pb collisions (Figure~\ref{fig:baryonmeson} and~\ref{fig:baryonmeson2}): for intermediate momenta (3--7 GeV/$c$) it exibits a relatively strong enhancement, by a factor of 3 compared to that in pp collisions at $p_{\rm T} \approx$ 3 GeV/$c$ and returns to the value in pp collisions at higher momenta ($p_{\rm T}$ above $\simeq$ 10 GeV/$c$)~\cite{OrtizVelasquez:2012te}. A similar observation was also reported for the $\rm \Lambda/K^{0}_{s}$ ratio and possible explanations have been proposed which include particle production via quark recombination~\cite{Fries:2003vb}.

\section{Transverse momentum distribution in proton-lead collisions}

Particle production in proton-lead (p--Pb) collisions allows one to study and understand QCD at low parton fractional momentum $x$ and high gluon density. It is moreover expected to be sensitive to nuclear effects in the initial state. For this reason p--Pb measurements provide an essential reference tool to discriminate between initial and final state effects and they are crucial for the studies and the understanding of deconfined matter created in nucleus-nucleus collisions.

\begin{figure}[t]
\centering
\includegraphics[width=0.7\linewidth,clip]{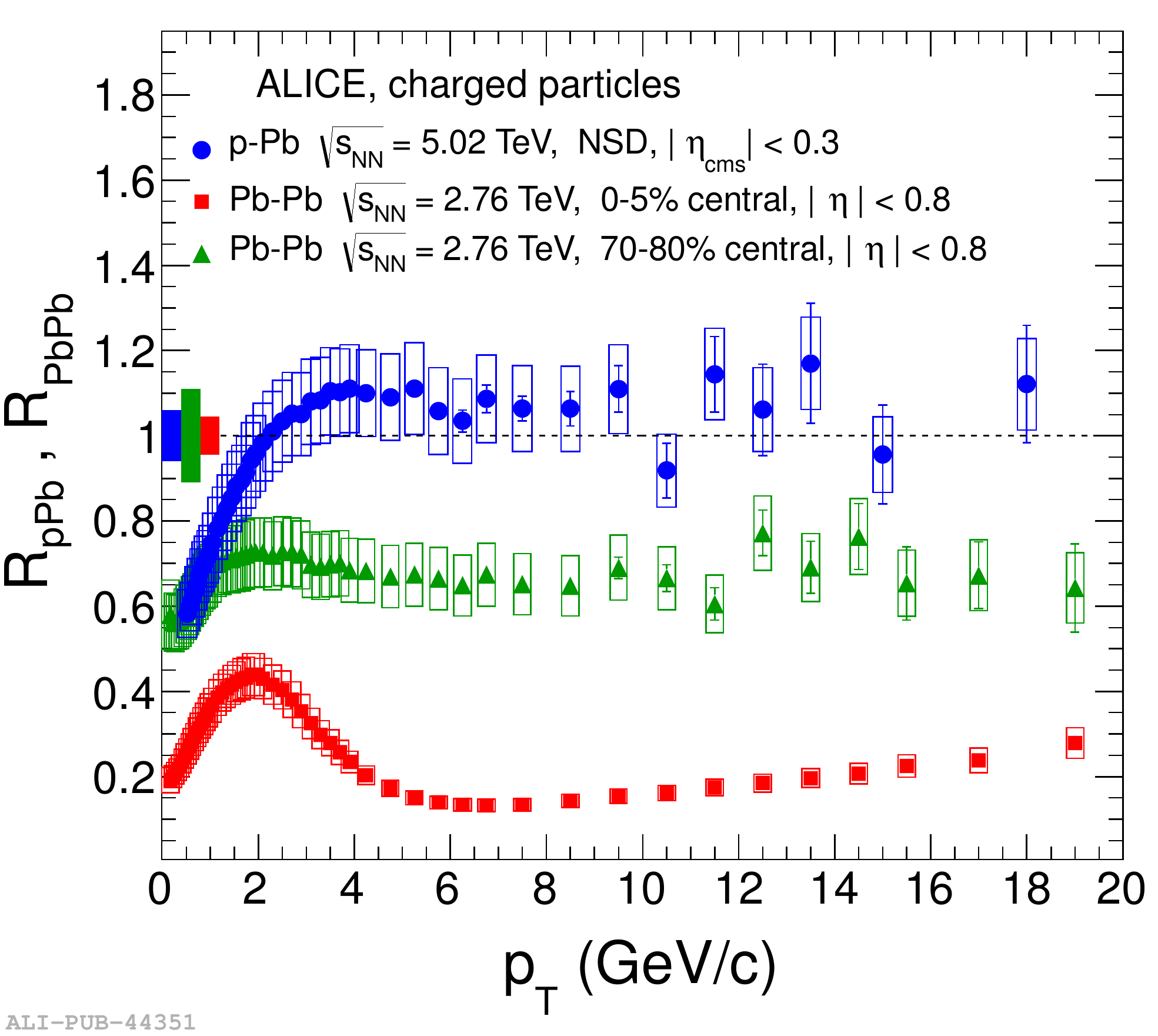}
\caption{The nuclear modification factor of charged particles as a function of transverse momentum in NSD p--Pb collisions at $\sqrt{s_{\rm NN}}$ = 5.02 TeV compared to measurements in central (0-5\%) and peripheral (70-80\%) Pb--Pb collisions at $\sqrt{s_{\rm NN}}$ = 2.76 TeV.}
\label{fig:rppb}
\end{figure}

The measurement of the transverse momentum $p_{\rm T}$ distributions of charged particle in p--Pb collisions were reported already in~\cite{ALICE:2012mj}. It was previously shown that the production of charged hadrons in central Pb--Pb collisions at the LHC is strongly suppressed~\cite{Aamodt:2010jd,:2012eq}. The suppression remains substantial up to 100 GeV/$c$ and is also seen in reconstructed jets~\cite{:2012is}. Proton-lead collisions provide a control experiment to clearly establish whether the initial state of the colliding nuclei plays a role in the observed high-$p_{\rm T}$ hadron production in Pb--Pb collisions. In order to quantify nuclear effects, the $p_{\rm T}$-differential yield relative to the proton-proton reference, the so-called nuclear modification factor, is calculated. The nuclear modification factor is expected to be unity for hard processes which exhibit binary collision scaling. This has been recently confirmed in Pb--Pb collisions at the LHC by the measurements of direct photon~\cite{Chatrchyan:2012vq}, Z$^{0}$~\cite{Chatrchyan:2011ua} and W$^{\pm}$~\cite{Chatrchyan:2012nt}, observables which are not affected by hot QCD matter. In Figure~\ref{fig:rppb} the measurement of the nuclear modification factor in p--Pb collisions R$_{\rm pPb}$ is compared to that in central (0-5\% centrality) and peripheral (70--80\%) Pb--Pb collisions R$_{\rm PbPb}$. R$_{\rm pPb}$ is observed to be consistent with unity for transverse momenta higher than about 2 GeV/$c$. This important measurement demonstrates that the strong suppression observed in central Pb--Pb collisions at the LHC is not due to an initial-state effect, but it is a final state effect related to the hot matter created in high-energy heavy-ion reactions. 

\section{Summary and conclusions}

ALICE has obtained so far a wealth of physics results both from the analysis of proton-proton collision data and from the first two LHC heavy-ion runs. 

The transverse momentum spectra of $\pi^{\pm}$, $K^{\pm}$, $p$ and $\bar{p}$ have
been measured with ALICE in several colliding systems and energies at the LHC
demonstrating the excellent PID capabilities of the
experiment. Data in pp collisions show no evident $\sqrt{s}$ dependence in
hadron production ratios. In Pb--Pb
collisions $\bar{p}/\pi^{-}$ integrated ratio is significantly 
lower than statistical model predictions with a chemical freeze-out temperature $T_{ch} \simeq 160-170$~MeV. The average transverse momenta and the transverse momentum spectra indicate a $\sim$10\% stronger radial flow than at RHIC energies.

In the intermediate transverse momentum $p_{\rm T}$ region, an enhancement of the baryon-to-meson ratio is observed. The maximum of the ratio is shifted to higher $p_{\rm T}$ with respect to RHIC measurements. The results of the measurements of charged pion, kaon, protons and antiproton production at high $p_{\rm T}$ were also presented. The nuclear modification factors for these species are similar in magnitude, suggesting that the medium does not significantly affect fragmentation.

First results from a short pilot run with proton-lead beams have been also reported. The measurement of the charged-particle transverse momentum $p_{\rm T}$ spectra and nuclear modification factor in p--Pb collisions at $\sqrt{s_{\rm NN}}$ = 5.02 TeV, covering 0.5 $< p_{\rm T} <$ 20 GeV/$c$, show a nuclear modification factor consistent with unity for $p_{\rm T} >$ 2 GeV/c. This measurement indicates that the strong suppression of hadron production at high $p_{\rm T}$ observed at the LHC in Pb--Pb collisions is not due to an initial-state effect, but is the fingerprint of jet quenching in hot QCD matter.

\end{document}